
\input harvmac
\parskip 6pt
\def\sqr#1#2{{\vcenter{\hrule height.#2pt \hbox{\vrule width.#2pt
height#1pt \kern#1pt \vrule width.#2pt} \hrule height.#2pt}}}

\def \l{\langle}
\def \r{\rangle}

\def \pr{\partial}

\def \d{{\rm d}}

\def \darrow#1{\raise1.5ex\hbox{$\leftarrow$}\mkern-16.5mu #1}
\def\darr#1{\raise1.5ex\hbox{$\leftrightarrow$}\mkern-16.5mu #1}

\lref\shorven{G. Shore and G. Veneziano, Nucl. Phys. {\bf B 381}
(1992) 23.}
\lref\thooft{G. 't Hooft, {\sl Recent developments in 
gauge theories}, Eds. G. `t Hooft {\sl et al.}, 
Plenum Press, New York, 1980.}
\lref\adler{S. L. Adler and W. Bardeen, Phys. Rev. {\bf 182} (1969) 1517.}
\lref\wilson{K. G. Wilson and J. Kogut, Phys. Rep. {\bf 12C} (1974) 75.}
\lref\zamref{A. B. Zamolodchikov, JETP Lett. {\bf 43} (1986)
730.}
\lref\cardy{J. L. Cardy, Phys. Lett. {\bf B215} (1988) 749.}
\lref\osbornone{H. Osborn, Phys. Lett. {\bf B222} (1989) 97\semi
    I. Jack and H. Osborn, Nucl. Phys. {\bf 343} (1990) 647.}
\lref\cfl{A. Cappelli, D. Friedan  and J.I. Latorre,
    Nucl. Phys. {\bf B352} (1991) 616.}
\lref\cfv{A. Cappelli, J. I. Latorre and X. Vilas\'\i s-Cardona,
    Nucl. Phys. {\bf B376} (1992) 510; hep-th/9109041.}
\lref\flv{D. Z. Freedman, J. I. Latorre and X. Vilas\'\i s-Cardona,
    Mod. Phys. Lett. {\bf A6} (1991) 531.}
\lref\osbornthree{J. I. Latorre and H. Osborn, Nucl. Phys. 
{\bf B511} (1998) 737; hep-th/9703196.}
\lref\osbornfreedman{H. Osborn and D. Z. Freedman,
Phys. Lett. {\bf B432} (1998) 353; hep-th/9804101.}
\lref\gomez{E. \'Alvarez and C. G\'omez, hep-th/9810102.}
\lref\zumbach{G. Zumbach, Nucl.Phys. {\bf B413} (1994) 754; Phys. Lett.
{\bf A190} (1994) 225\semi
   J. Generowicz, C. Harvey-Fros, T. R. Morris,
Phys. Lett. {\bf B407} (1997) 27; hep-th/9705088\semi
P. Haagensen {\sl et al.}, Phys. Lett. {\bf B323} (1994) 330;
hep-th/9310032.}
\lref\bastianelli{F. Bastianelli, Phys. Lett. {\bf B369} (1996) 249;
hep-th/9511065.}
\lref\freedman{D. Anselmi, J. Erlich, D. Z. Freedman, A. Johansen;
hep-th/9711035\semi
D. Anselmi, D. Z. Freedman, M. T. Grisaru, A. A. Johansen;
hep-th/9708042.}
\lref\seiberg{
N. Seiberg, Phys. Rev {\bf D49} (1994) 6857;  hep-th/9402044\semi
N. Seiberg, Phys. Lett {\bf B435} (1995) 129; hep-th/9411149\semi
K. Intriligator and N. Seiberg,
Nucl. Phys. Proc. Suppl. {\bf 45BC} (1996) 1; hep-th/9509066.}
\lref\shore{G. M. Shore, Phys. Lett. {\bf B253}
(1991) 380; {\bf B256} (1991) 407.}
\lref\polyakov{A. Polyakov, Phys. Lett. {\bf B103} (1981) 207.}
\lref\wiconf{
C. Callan, S. Coleman and R. Jackiw, Ann. of Phys. {\bf 59}
(1970) 42}
\lref\wirev{S. Coleman, {\it Aspects of Symmetry},
(Cambridge, Cambridge, U.K., 1985)\semi
R.~Jackiw, in S.~B.~Treiman {\sl et al.}, 
{\it Current Algebra and Anomalies} (World Scientific,
Singapore, 1985).} 
\lref\widet{J. C. Collins, {\it
Renormalization}, (Cambridge, Cambridge, U.K., 1984).}
\lref\scwir{J. C. Collins, A. Duncan and S. D. Joglekar, {\it
Phys. Rev.} {\bf D16} (1977) 438\semi N. K. Nielsen, {\it Nucl. Phys.}
{\bf B120} (1977) 212\semi R. Tarrach, {\it Nucl. Phys.} {\bf B196}
(1982) 45.}
\lref\duff{M. J. Duff, Nucl. Phys. {\bf B125} (1977) 334.}
\lref\birrell{N. D. Birrell and P. C. W. Davies, 
{\sl Quantum fields in curved space}, Cambridge Univ. Press (1982).}
\lref\fl{S. Forte and J. I. Latorre,
 hep-th/9805015, {\bf B511} 
(1998) 737.}
\lref\knecht{M. Knecht and E. de Rafael, Phys.Lett. {\bf B424} (1998) 335;
 hep-ph/9712457\semi
M. Knecht, S. Peris and E. de Rafael, hep-ph/9809594.}
\lref\witten{E. Witten, Phys. Rev. Lett. {\bf 51} (1983) 2351.}
\lref\comellas{J. Comellas, J. I. Latorre and J. Taron,
Phys, Lett. {\bf B360} (1995) 109; hep-ph/9507258. } 
\lref\holzhey{C. Holzhey, F. Larsen and F. Wilzek, Nucl. Phys. {\bf B424}
(1994) 443; hep-th/9403108.}
\lref\gaite{J. Gaite and D. O'Connor, Phys. Rev {\bf D54} (1996) 5163;
hep-th/9511090\semi
J. Gaite, hep-th/9810107.}
\lref\fradkin{A.H. Castro Neto
 and E. Fradkin, Nucl.Phys. {\bf B400} (1993) 525;  cond-mat/9301009.}
\lref\specan{D. Boyanovsky and R. Blankenbecler, Phys. Rev. {\bf D31}
(1985) 3234.}

\noblackbox
\pageno=0\nopagenumbers\tolerance=10000\hfuzz=5pt
\baselineskip=12pt plus2pt minus2pt
\newskip\footskip\footskip=10pt 
\rightline{hep-th/9811121}
\rightline{UB-ECM-PF 98/24}
\rightline{INFN-RM3 98/8}
\rightline{DFTT 65/98}
\vskip 10pt
\centerline {\bf Realization of symmetries and the $c$--theorem}
\vskip 24pt
\centerline {Stefano Forte${}^a$\footnote*{On leave from INFN, 
Sezione di Torino, Italy} and Jos\'e I. Latorre${}^b$}
\vskip 12pt
\centerline {${}^a$\it I.N.F.N., Sezione di Roma III}
\centerline {\it Via della Vasca Navale 84, I-00146 Rome, Italy}
\vskip 10pt
\centerline{${}^b$\it 
Departament d'Estructura i Constituents de la Mat\`eria,}
\centerline{\it Universitat de Barcelona}
\centerline{and {\it  I.F.A.E.,}}
\centerline {\it Diagonal 647, E-08028 Barcelona, Spain}
\vskip 40pt

{\centerline{\bf Abstract}
\medskip\narrower
\baselineskip=10pt

We discuss the relation between the $c$--theorem and the
 the way various symmetries are realized in quantum field theory.
We review our recent proof of the $c$--theorem in four dimensions.
Based on this proof and further evidence,
 we conjecture that the realization of chiral
symmetry be irreversible, flowing from
the Wigner-Weyl realization at short distance to 
the Nambu--Goldstone realization 
at long distance. We argue that three apparently independent
 constraints based on the renormalization group, namely
 anomaly matching, the
$c$--theorem and the conjectured symmetry realization theorem, are
particular manifestations of a single underlying principle.  
\smallskip}
\vskip 48pt

\centerline{ Presented by J.I.L. at the }
\centerline{Faro Workshop on Exact Renormalization Group}
\centerline{ September 10-12, 1998.}
\medskip
\centerline{\it to be published in the proceedings
}
\vskip 64pt
\line{November 1998\hfill}

\vfill\eject \footline={\hss\tenrm\folio\hss}

\newsec{Introduction}

The $c$--theorem expresses a  deep property of the 
renormalization group (RG), namely the irreversibility of its
flows for unitary quantum field theories. This irreversibility is
proven by explicitly exhibiting an observable function ($c$--function) 
which is
monotonically decreasing along RG trajectories.
The theorem, originally 
discovered~\zamref\ in a two-dimensional setting, 
has been subsequently 
analyzed from a variety of
 viewpoints~\refs{\cardy\osbornone\cfl\shore\fradkin\gaite\cfv\flv\osbornthree
-\gomez}, and recently generalized to four dimensions~\fl,
along a suggestion of ref.~\cardy,
by  means of  techniques which encompass
much of the previous effort. 

This recent result leads naturally
to examining the way symmetries are realized
along RG trajectories. 
Indeed, the decrease, proven in ref.~\fl, 
of the $c$--function
proposed in ref.~\cardy\  is supported by several explicit examples,
where it is a consequence of the fact that massless 
degrees of freedom tend to be bosonic in the infrared limit. 
This suggests that 
the $c$--theorem may be related to the way chiral symmetry is realized 
along the RG trajectory, i.e. that the RG  takes from Wigner-Weyl
to Nambu-Goldstone realizations when flowing from high to low energy.

In the sequel, we will formulate more precisely this idea, and  
conjecture that  chiral symmetry realization  be similarly governed by
an irreversibility theorem, and thus 
constrained by general principles such as unitarity. 
More in general, we will suggest that 
the $c$--theorem is
a particular case of
an  underlying algebra of scale and axial currents, whose Ward
identities govern the ways various symmetries
are realized. Anomaly matching, the $c$--theorem, the Goldstone
theorem and its flow are then different manifestations of a
single 
 deeper principle. The physical mechanism 
ultimately responsible for such a remarkable fact
appears to  be   the decoupling
of positive-norm massive states from the Hilbert space as one flows
towards the infrared. Different constraints are then
derived, depending on the Ward identity which is being
considered, and thus on the relevant intermediate states.

A  proof of this conjecture may provide a significant step in relating
renormalization group ideas to the concept of entropy in quantum 
field theory.

\newsec{Proving the $c$--theorem}

In a seminal paper \zamref, 
Zamolodchikov proved that an observable quantity
$c\left(g^i,\mu\right)$ based
on correlators of the energy--momentum tensor 
of a two dimensional, Poincar\'e invariant, 
renormalizable and unitary theory obeys an
irreversible renormalization group equation
\eqn\zam{-\beta^i {\pr\over \pr g^i} c(g^j,\mu) 
 \leq 0\ ,}
where $\mu$ is the renormalization scale (subtraction point).
It is then clear that RG flows are irreversible since
eq.~\zam\ entails an 
ordering of  theories along scales. Loosely speaking,
the theorem states that a net loss of information is
taking place when we move towards the infrared.

Though we will not enter the details
of the proof, it is  worth emphasizing some of its  key
features:
\item{ i)} The $c$--function
 is constructed from the energy-momentum tensor. This is
the only operator
in a quantum field theory which is guaranteed to exist and
 whose properties are always known 
since it is a descendant of the identity. Furthermore,
its anomalous dimension vanishes, thus simplifying the 
analysis.
\item{ ii)} The $c$--function is additive: the contributions to it
from two 
sets of non-interacting degrees of freedom
simply add up. This property suggests that the $c$--function
counts massless effective degrees of freedom in the Hilbert space.
\item{ iii)} The proof is based on  trading two
scales. The variation of energy--momentum tensor correlators
upon a change of $\mu$ (the subtraction point) is 
traded with  their flow upon variation of  $x$ (the space 
separation in a two--point energy--momentum tensor
correlator). Ward
identities for the conservation of the energy--momentum tensor
bring information on the change in $x$ and turn to be
sufficient to prove that   the variation of $c$ is proportional to
the correlator of two traces of the energy--momentum tensor.
\item{ iv)} Unitarity is at the origin of irreversibility. 

\noindent
The treatment of
the two-dimensional case is dramatically simplified by
the absence of spin structures in correlators of energy--momentum
tensors. In more than two dimensions, the role of 
the change of scale generator is taken by spin 0, whereas
the form of Green function at finite distances in conformal
field theories is governed by spin 2.

The two--dimensional proof of the $c$--theorem can 
be recast in terms of 
spectral densities~\cfl, making
apparent the role of positive--norm intermediate states, and thus
of unitarity which guarantees this positivity.
In this approach, the $c$--function is a measure of the 
massless intermediate
degrees of freedom in the correlator of two energy--momentum tensors,
and 
its decrease is due to the decoupling of modes as the scale is
lowered.

The road to the generalization of the $c$--theorem 
to higher dimensions
was initiated by Cardy~\cardy,
then pursued by a number of 
groups \refs{\osbornone,\cfl,\shore}, on the basis of
the observation that the central
charge of a $d=2$
conformal field theory is responsible for both the coefficient of the
two-point correlator of the energy-momentum tensors and for the
trace anomaly $\theta$ on a curved background.
 In general, in $d=2 n$ dimensions,
the trace anomaly appears both  in $n+1$--vertex graphs, and 
in the one-point function $\langle \theta\rangle$. 
The RG flow of these Green functions can then again be studied by
trading scales.  In two dimensions,
we may trade the subtraction point with the space separation of 
the two--point 
correlator, or with
the curvature scale of the one-point function.
In four dimensions,  one possibility is to work
with three-point functions; in order to
keep the analysis manageable one could then choose a
symmetric geometry and deal with a single overall 
scale factor. The alternative, easier option is
to study the  variation of the
curvature scale in the one-point function. The argument has then the
additional advantage of naturally generalizing to arbitrary even
dimension.

The proof of the $c$--theorem of ref.~\fl\ is  based on this line of
thought \cardy\cfl: one starts with 
a slight modification of Cardy's proposal 
for a $c$--function, namely the
vacuum-expectation value of the trace of
the energy--momentum tensor on a hyperbolic background
with constant curvature $R=-a^2 d(d-1)$:
\eqn\candidate{c\left({\mu\over a},g^i\right)
 \equiv {a^{-d} V\over A_d} \l\theta\r\ ,}
where $V$ is the volume of the $(d-1)$--dimensional sphere and
$A_d$ is a suitable  normalization factor.
This $c$--function obeys a renormalization group equation with
vanishing anomalous dimension:
\eqn\cequation{\mu{\d\over \d\mu}c\left({\mu\over a},g^i\right)=
\left(\mu{\partial\over \partial\mu} +
\beta^i(g){\pr\over \pr g^i}\right) c\left({\mu\over a},g^i\right) =0\ .}
The dependence on the subtraction point is then traded for the
dependence on the curvature $a$, whose variations are generated by
\eqn\scaltr{\delta_s\equiv a{\d\over \d a}= -{2}\int \d^dx\  g^{\alpha\beta}(x)
{\delta\over \delta g^{\alpha\beta}(x)}\ .}

The effect of a scale transformation $\delta_s$ is
dictated by the scale Ward identity~\refs{\wiconf-\scwir}
\eqn\genwi{-{2}g^{\alpha\beta}(x)
{\delta\over \delta g^{\alpha\beta}(x)} \langle{\cal O}(y)\rangle=-
\nabla_\mu\l j^\mu_D(x)
{\cal O}(y)\r+\delta^{(d)}(x-y)
\langle\delta_s{\cal O}(y)\rangle+\langle
\theta(x)  {\cal O}(y)\rangle\ ,}
where only the last term on the r.h.s., namely
the insertion of the
trace of the energy-momentum tensor, is a non-contact contribution.
Note that only contributions to $\theta$ proportional to
the beta functions can contribute to this term unless $\cal O$ is
proportional to the identity operator, because the remaining
contributions to $\theta$ are proportional to the identity, and the
correlator only includes connected contributions.

Identifying the operator $\cal O$ with that which appears in
the definition of the $c$--function eq.~\candidate, all contact terms
cancel in eq.~\genwi, and we are led to
\eqn\finalth{-\beta^i \pr_i c = -{1\over V}\int \! \d^dx\ 
\sqrt{g(x)} \l\theta_{dyn}(x)\theta_{dyn}(0)\r_s \leq 0\ ,}
where $\theta_{dyn}$ is the sum of the
contributions to the trace of the
energy--momentum tensor which are proportional
to the beta-functions
 and the subscript  $s$ indicates that the correlator is free of
contact  terms. The irreversibility of the flow follows from 
the positivity of the correlator on
r.h.s. of eq.~\finalth, which in turn can be proven 
for unitary theories by means of a
spectral representation.

The physical interpretation of the $c$--theorem is 
conceptually similar to that of its two-dimensional incarnation  discussed
previously: 
massive, positive--norm intermediate states in the
correlator $\langle \theta\theta\rangle$ decouple in the infrared,
and this
produces irreversibility of RG flows. 
The proof is 
done at the level of Hilbert space states, so in the above statement
no reference is implied to
the  fields which define the ultraviolet theory.

The main subtlety in the proof is related to the absence of
contact terms on the r.h.s. of eq.~\finalth, which is crucial in
establishing that its sign is negative semi-definite.
This is closely related to the issue of renormalization scheme 
dependence. Indeed,
when perfoming a change of subtraction scheme, two modifications
take place: the correlator of the two energy--momentum tensors
picks a contact term proportional to $\langle \theta\rangle$
and the Ward identity acquires a contribution which
exactly cancels it. The physical reason for this
effect is the fact that a change of scale of $c$ is observable
and, thus, scheme independent. The above 
proof was presented in 
the particular scheme in which the Ward indentities take the form of
eq.~\genwi, i.e. such that the equations of motion are satisfied at
the operator level by the renormalized operators.

This proof of the $c$--theorem, however, has some shortcomings.
The most annoying one is
the need to work in curved space. A more natural proof
should only use properties of flat space; however it is then very
likely that it should make use of 
three-point functions of the energy--momentum tensor.
It should also be possible to get a Wilsonian
proof of the $c$--theorem along the lines initiated in
ref.~\zumbach. This would extend the validity of the theorem to
effective theories, even if not renormalizable:
for instance, chiral perturbation theory looks
irreversible though not renormalizable.   
At a deeper level, one would like to understand the $c$--theorem in
terms of information theory, i.e. relate the $c$--function and its
decrease to a characterization of the entropy of the system \gaite.
 We shall come back to this last
point in our conclusions.

\newsec{Realizations of  the $c$--theorem}

On top of the formal proof~\fl, a considerable amount of direct
evidence
supports the conclusion that Cardy's $c$--function eq.~\candidate\ 
decreases along RG flows. 
In fact, it was already proven in ref.~\cardy\ that
the $c$--function decreases
to first non--trivial order in conformal perturbation theory. 
Notice that this
includes all the flows around any fixed point, gaussian or not.
Furthermore, the  inequality $c_{UV}-c_{IR}\ge 0$ can be checked
explicitly in several cases. 

A striking instance  is in
supersymmetric gauge theories, where the desired inequality is
fulfilled by an overwhelming number of exactly known
realizations~\refs{\bastianelli,\freedman,\osbornfreedman}. 
A representative example is the ${\cal N}=1$ supersymmetric
Yang-Mills theory with $SU(N_c)$ color gauge group
and $N_f$ flavors in the fundamental representation. 
The ultraviolet $c$--charge is~\bastianelli
\eqn\cuv{c_{UV}=(N_c^2-1) c_V + 2 N_f N_c c_S\ ,}
where the charges for a vector and a scalar supermultiplet
are computed by adding those of their components and found to
be $c_V=135$ and $c_S=15$.
According to
the exact solution of ref.~\seiberg, we get 
\medskip
\settabs\+aaaa&aaaaaaaaaaaaaaaaaaaaaaaa&aaaaaaaaaaaaaaaaaaaaaaaa&\cr
\+&$N_f=0$&$c_{IR}=0$\ \ \ (massgap)\cr
\+&$0<N_f<N_c$& no vacuum\cr
\+&$N_f=N_c$&$c_{IR}=\left(N_f^2+1\right) c_S$\cr
\+&$N_f=N_c+1$&$c_{IR}=\left(N_f^2+2 N_f\right) c_S$\cr
\+&$N_c+2\leq N_f< {3\over 2} N_c$&$
c_{IR}=\left((N_f-N_c)^2-1\right) c_V + \left(
(3 N_f-2 N_c) N_f\right) c_S$\cr
\+&${3\over 2}N_c=N_f$&$
c_{IR}=\left({N_c^2\over 4}-1\right)+{15\over 4} N_c^2 c_S$\cr

\medskip\noindent
So  $c_{UV} < c_{IR}$ always. 
The same test can be performed with
gauge groups $SO(N_c)$, $Sp(2 N_c)$, $G_2$ and choosing
different representations for the matter fields. 
The analysis can also be carried
through for theories with non-trivial infrared realizations in a
systematic and exhaustive way~\freedman, and its result likewise
supports the theorem.

A particular case of obvious interest is that of QCD, already
discussed by Cardy~\cardy. Here, the ultraviolet degrees
of freedom are free fermions and vectors, whose $c$--charges add up to  
\eqn\qcduv{c_{UV}= \left(N_c^2-1\right) c_v + 2 N_f N_c c_f}
with $c_v=62$ and $c_f=11$.
The standard infrared realization in the chiral limit, with
spontaneously broken chiral symmetry, gives
\eqn\qcdir{c_{IR}= \left(N_f^2-1\right) c_s\ ,} 
with $c_s=1$ and does obey $c_{IR}<c_{UV}$.
Notice that a Wigner-like realization
of chiral symmetry 
with $2 N_f^3$ massless baryons would instead violate the $c$ theorem:
its $c$--charge would be larger than the ultraviolet one.

Note furthermore that the r.h.s. in eq.~\finalth\ is
proportional to the square of the beta--function. It is thus easy to
see
 that in the large $N_c$
limit the derivative of $c$ contain terms of the form
\eqn\largen{\dot c\sim\ \langle \theta \theta\rangle \sim 
a N_c^2 + b N_c N_f + c N_f^2 \ .}
The first two types of terms have the structure needed to remove 
the corresponding contributions to
$c_{UV}$ whereas the last leads to the value of $c_{IR}$
associated to the pions. It is  curious to observe that the 
factor 11 
which appears in the beta--function matches the value of 
the  $c$--charge coming from the trace anomaly. 
The $c$--theorem might provide a connection between
these coefficients. 

%

\newsec{The trace anomaly and the axial anomaly }

The $c$--theorem, which provides a constraint on short-- {\sl vs.}
long--distance realizations of quantum field theories, appears to be
intimately connected to the trace anomaly.
 A different set
of constraints, the 't~Hooft anomaly matching conditions, are instead
related to the axial anomaly.
Axial and trace
anomalies are characterized at fixed points
by coefficients (charges)
 that multiply topological terms. The behavior of these charges under RG 
transformations is not the same: axial anomalies are protected by the 
Adler-Bardeen non-renormalization theorem and thus scale--independent,
 whereas the trace anomaly
scales in an irreversible way. There is however a close kinship
between axial and conformal symmetries and the respective anomalies.

Indeed, consider the chiral Ward identity~\wirev\
associated to the chiral
variation $\delta_5\psi= i\gamma_5 \psi$ of the fermion fields 
$\psi$: this has the same
form as the conformal Ward identity eq.~\genwi, except
that  the
l.h.s. (which is
due to transformation of the space--time coordinates) 
vanishes since the axial symmetry is an internal symmetry:
\eqn\chirwi{0=-
\nabla_\mu\l j^\mu_5(x)
{\cal O}(y)\r+\delta^{(d)}(x-y)
\langle\delta_5{\cal O}(y)\rangle+\langle
\partial_\mu j^\mu_5  {\cal O}(y)\rangle\ .}
The last term on the r.h.s. of eq.~\chirwi\ is the sum of a classical
chiral 
symmetry breaking term
(proportional to the fermion mass $m_\psi$) and an axial anomaly ${\cal A}(x)$:
\eqn\aneq{
\partial_\mu j^\mu_5= 2i m_\psi\bar\psi\gamma_5\psi+{\cal A}.}

Now, it is interesting to consider various possibilities 
for the way chiral symmetry is realized; these can be studied by
considering various choices for the operator $\cal O$ in 
eq.~\chirwi.  First, consider the case in which the axial symmetry is
broken by an anomaly, and take $\cal O$ proportional to 
the identity operator. In the absence of massless excitations we are
then left with
\eqn\indth{0= 2i\l m_\psi\bar\psi\gamma_5\psi\r+\l{\cal A}\r.}
As is well known~\wirev, the first 
term on the r.h.s. of eq.~\indth\ is
equal to (twice) the number of right-handed minus left-handed
zero-modes of the Dirac operator: the zero modes are 
thus counted by the
anomaly, which must match them exactly in order to produce the desired
cancellation in eq.~\indth\ (index theorem). 

Furthermore, we may study the scale dependence of either of the two
terms on the r.h.s. of eq.~\indth\ by taking the operator $\cal O$ in
the conformal Ward identity~\genwi\ proportional to it:
if we take in particular  
\eqn\chirop{{\cal O}={\cal O}_5
\equiv2iN m_\psi\bar\psi\gamma_5\psi,} 
with $N$ a normalization
factor chosen in such a way that $\cal O$ is dimensionless, we get
\eqn\wianomaly{\delta_{scale} \l N{\cal A}\r=-
\delta_{scale} {\cal O}_5
=-\int \langle \left[ 2iN m_\psi\bar\psi\gamma_5\psi\right]
  \theta \rangle=0. }
The last step follows from the fact that there are no
 intermediate states that couple to both $\theta$ and $\bar\psi
\gamma_5\psi$, as it is apparent recalling~\scwir\ 
that the fermion
contribution to $\theta$ is proportional to $\bar\psi \psi$.
It follows that there is no flow for the anomaly: the axial anomaly
must exactly match in high-- and low--energy theories.

The relation between the RG flows of the scale and axial anomalies
can be pursued at the
level of spectral representations. Whereas the monotonic nature of the
former flow can be related to the decoupling of massless modes 
towards
the infrared, the lack of flow of the latter is reflected in the fact
that the spectral representation for the axial anomaly satisfies a sum
rule which relates its infrared and ultraviolet behaviors~\specan.
Indeed, the anomaly $\cal A$ can be computed from the ultraviolet
behavior of the fermion spectrum, whereas the index theorem
eq.~\indth\ relates it to the number of zero modes. Since for
supersymmetric theories the axial and conformal current belong to the
same supermultiplet, it is suggestive to conjecture that both the
anomaly matching and the $c$--theorem may be manifestations of the
algebra of this supercurrent, whereby commuting the scale current with
either the scale or chiral current generates scale transformations of
the respective anomalies.

The set of algebraic conditions is then completed by studying
the commutator of the axial current with itself. This corresponds to
considering chiral variations of a chiral operator such as ${\cal
O}_5$ eq.~\chirop. The ensuing chiral Ward identities are rather
complex in the general case in which both the anomaly and a fermion
mass are present (see for instance ref.~\shorven\ for recent
developents), but they reduce to a simple and well known case
in the chiral limit and in the absence of axial anomaly
(such as
when the current $j^\mu_5$ is flavor non-singlet):  
\eqn\golth{0=-
\nabla_\mu\l j^\mu_5(x)
{\cal O}_5(y)\r+\delta^{(d)}(x-y)
\langle\delta_5{\cal O}_5(y)\rangle ,}
where we are assuming that the normalization factor in eq.~\chirop\
is now
chosen in such a way that ${\cal O}_5$ admits a finite chiral limit.
Eq.~\golth\ is just
the Ward identity that leads to Goldstone's theorem~\wirev.
Indeed, it can be satisfied in two ways:
either the two terms on the r.h.s. of eq.~\golth\ separately
vanish, or they are equal  and opposite. In the latter case
the vacuum breaks chiral symmetry, and 
there must exist massless particles with the quantum
numbers of ${\cal O}_5$. So the two options correspond to chiral 
symmetry
being realized in the Wigner or Goldstone mode respectively.
This now suggests that one may further pursue the study of the 
algebra
of chiral and scale currents by looking at the scale dependence of
Goldstone's theorem. We will do this in the next section.

\newsec{Irreversibility of the realization of chiral symmetry}

Most of the examples presented as evidence for 
the $c$--theorem based on Cardy's $c$--function share
the  remarkable property of displaying a Goldstone realization 
of some flavor symmetry at long distances.
Indeed, the checks are
 performed verifying
that $c_{UV}\le c_{IR}$: this is possible because 
the realizations of the flavor symmetry are known at 
both endpoints of the RG flow, and turn out to
be free theories at those points. For instance, QCD is 
a theory of
free massless quarks and gluons (strictly non-interacting
right at the fixed-point) in the UV and a theory of free 
massless pions (in the chiral limit) in the IR.

The key observation is then 
that free massless bosons, fermions and vectors weight,
according to the trace anomaly, as 
\eqn\cbfs{c_{b}=1\qquad c_{f}=11\qquad c_{v}=62 \ .}
These numbers encode some deep, basic property of
quantum field theory. For instance, considering the above values
 together with those  
given by the coefficient proportional 
to the Weyl tensor in the trace anomaly  
(1 for bosons, 6 for fermions and 12 for
vectors) immediately
implies the impossibility of a trivial bosonization in $d=4$.
Moreover, massless bosons are favored in the infrared in the
sense that their weight in the $c$--charge is smaller
than the ones of fermions and vectors:
the $c$--theorem, then, generates a constraint on
which infrared massless modes may  describe the system, and in
particular favours the Goldstone 
realization of chiral symmetry over the 
Wigner-Weyl realization. Indeed, there are no known examples in which
once a Nambu--Goldstone realization is obtained, a
Wigner-Weyl realization is recovered at lower energy.

This leads to conjecture a property of irreducibility for chiral symmetry
realizations. Consider for 
example the case of QCD. As we have seen in the end of the last
section, whether chiral symmetry is realized in the Goldstone
or Wigner mode hinges on whether the two terms on the r.h.s. of
eq.~\golth\ vanish or not. 
In order to study the scale dependence of symmetry
realization it is thus
sufficient to study the scale dependence of 
the
expectation value $\l {\cal O}_c\r$
of the chiral order parameter 
\eqn\chirpar{{\cal O}_c\equiv \delta_5
{\cal
O}_5.}  
But this is just given by 
the Ward identity eq.~\genwi:
if the order parameter is
dimensionless and observable, from it follows 
the analogue of eq.~\finalth, namely
\eqn\conjth{-\beta^i \pr_i \langle {\cal O}_c\rangle 
= -{1\over V}\int \! \d^dx\ 
\sqrt{g(x)} \l\theta_{dyn}(x){\cal O}_c\r_s  .}
If the correlator on the r.h.s. of eq.~\conjth\
can   be
proven to be positive definite, then
the realization of chiral symmetry is
irreversible:
 if the order parameter moves away from
zero it can never go back to zero. This means that as soon as chiral
symmetry is spontaneously broken, it remains 
broken at any lower scale. Irreversibility is once again  
related
to the decoupling of positive--norm (unitarity) intermediate 
states, now from the r.h.s. of eq.~\conjth.

A formal proof of 
this result presents several technical complications. The most 
notable is related to the way the chiral limit is approached:
either we stay away from the limit, in which case we must deal with
a more complicated form of the chiral Ward identity, or we are at the
limit, in which case the definition of the operator \chirop\ is
nontrivial. Also, even though the operator ${\cal O}_c$ is
proportional to the fermionic contribution to $\theta$, the 
correlator on
the r.h.s. of eq.~\conjth\ is not manifestly a square, and thus
it is not manifestly positive. Therefore, for the time being
we present the irreversibilty of the realization of 
chiral symmetry 
as a conjecture, which we expect to hold for some choice of the chiral
order parameter ${\cal O}_c$, not necessarily identical to that of
eq.~\chirpar-\chirop. 


Indeed, an 
alternative option  is to choose a many--point function as a
chiral order parameter: in such case we can then dispense with the
need to deal with curved space, since we may use the
 momentum transfer
as a scale. A natural candidate is based on a VV-AA 
correlator:\refs{\witten-\knecht} 
\eqn\pimunu{
\Pi_{LR}^{\mu\nu}(q)\equiv 2 {\rm i} \int d^4x\
{\rm e}^{{\rm i} q\dot x}\ \l 0\vert 
T\left( L^\mu(x)R^\nu(0)^\dagger\right)\vert 0\r \ ,}
where
\eqn\defrl{R^\mu,L^\mu= \bar d \gamma^\mu {1\over 2}
\left( 1\pm \gamma_5\right) u(x)\ . }
Extracting the Lorentz structure according to
\eqn\defpi{\Pi_{LR}^{\mu\nu}(q)\equiv\left( q^\mu q^\nu -g^{\mu\nu}
q^2\right) \Pi_{LR}\left(Q^2\right) \ ,}
with $Q^2=-q^2$, we may take as a chiral order parameter
the function $F(Q^2)$ defined by
\eqn\deff{F\left(Q^2\right) \equiv -Q^2 \Pi_{LR}\left(Q^2\right).}
Indeed, $F(Q^2)$ 
is zero in perturbation theory,
while at long distances chiral perturbation
theory gives
\eqn\flong{F\left(Q^2\right) = f_\pi^2 + 4 L_{10} Q^2 +\dots .}
The RG flow  $F(Q^2)$ can then be studied by trading its scale dependence
 for the dependence on its argument $Q^2$. 
Available knowledge on $F(Q^2)$~\refs{\witten,\knecht}\ seems to
support its irreversibility; 
a formal proof, however is nontrivial. It is
interesting to observed that  if the irreversibility were proven, then
 the ordering of
resonances in the large $N_c$  limit would be strongly
 constrained~\knecht.

\newsec{ RG flows and information theory}

The ultimate understanding of irreversibility of RG flows must
be rooted in information theory. This, however, is only
well--understood at the classical level, 
where Shannon's entropy provides
a quantitative measure of our lack of information of a system.
Its quantum counterpart is based on von Neumann's entropy
associated to a density matrix.
One feature of entropy in quantum mechanics which appears to be 
relevant
 here is the fact that, given a system with several subsystems,
entropy increases upon subtracing over one of them.  
This leads directly to the renormalization group, for
a RG
transformation can be viewed as a
trace over degrees of freedom which encompasses a loss
of information.

However, promoting  von Neumann's ideas to quantum field theory
is not obvious~\gaite. To begin with, several definitions of entropy are
available, not all of which are simply related  
to the original quantum
mechanical one. 
One of these definitions, namely the geometric entropy~\holzhey, 
does have the
property of increasing upon subtracing a system. 
In two dimensions
the change of this entropy with respect to the variation of
an infrared scale can be
shown to be related to the trace of 
the energy--momentum tensor and thus to the $c$--charge. 
It is tempting to guess that such a description might provide a
deeper understanding of the $c$--theorem.

\bigskip
\goodbreak
\noindent{\bf Acknowledgments}
\nobreak
J.I.L. thanks M. Asorey,
  M. Knecht, 
C. Korthas-Altes, C. A. L\"utken,   H. Osborn, E. de Rafael, 
 and G. Shore for many fruitful
discussions.
Financial support from 
CIRIT (contract 1996GR00066) is also acknowledged.

\listrefs
\bye